\documentclass[5p]{elsarticle}
\usepackage{graphicx}
\usepackage{amsmath}
\usepackage{color}

\def\bra{\langle}
\def\ket{\rangle}
\def\rmd{{\rm d}}

\def\half#1{{#1\over 2}}

\def\lamilamj{{\lambda_i\cdot\lambda_j}}

\def\sigisigj{{\sigma_i\cdot\sigma_j}}

\def\calA{{\cal A}}

\def\calP{{\cal P}}

\def\calPcs{{\cal P}}

\def\crsth{{[$q^3$8\half3]}}
\def\crsh{{[$q^3$8\half1]}}
\def\cwsh{{[$q^3$1\half1]}}

\def\Lamc{\mbox{$\Lambda_c$}}
\def\Sigc{\mbox{$\Sigma_c$}}
\def\Sigcstar{\mbox{$\Sigma_c^*$}}

\def\Jpsi{\mbox{$J\!/\!\psi$}}
\def\etac{\mbox{$\eta_c$}}
\def\Dbar{\mbox{$\overline{D}$}}
\def\Dbarstar{\mbox{$\overline{D}{}^*$}}

\def\cbar{\overline{{c}}}
\def\bbar{\overline{{b}}}
\def\qbar{\overline{{q}}}
\def\Qbar{\overline{{Q}}}

\newcommand{\xbld}[1]{\mbox{\boldmath $#1$}}

\def\vecr{{\xbld{r}}}

\def\vecp{{\xbld{p}}}

\def\vecP{{\xbld{P}}}

\def\vecR{{\xbld{R}}}

\def\vecSp{{\xbld{S}}}

\def\rmd{{\rm d}}

\def\half#1{\text{${#1\over 2}$}}

\def\Vcoul{V_\text{Coul}}
\def\Vconf{V_\text{conf}}
\def\Vcmi{V_\text{CMI}}
\def\Ocmi{{{\cal O}_\text{CMI}}}

\begin{document}

\begin{frontmatter}

\title{The hidden charm pentaquarks are the hidden color-octet $uud$ baryons?}
%\author{Sachiko Takeuchi\corref{cor1}$^{1,3,4}$ and Makoto Takizawa$^{2,4,5}$}
%\address{$^1$Japan College of Social Work, Kiyose, Tokyo 204-8555, Japan\\
%$^2$Showa Pharmaceutical University, Machida, Tokyo 194-8543, Japan\\
%$^3$Research Center for Nuclear Physics (RCNP), Osaka University, Ibaraki, Osaka, 567-0047, Japan\\
%$^4$Theoretical Research Division, Nishina Center, RIKEN, Hirosawa, Wako, Saitama 351-0198, Japan\\
%$^5$J-PARC Branch, KEK Theory Center, Institute of Particle and Nuclear Studies, KEK, Tokai, Ibaraki, 319-1106, Japan
%}
\author[JCSW,RCNP,RIKEN]{Sachiko Takeuchi\corref{cor1}}
\ead{s.takeuchi@jcsw.ac.jp}
\author[Showa,RIKEN,JPARC]{Makoto Takizawa}

\address[JCSW]{Japan College of Social Work, Kiyose, Tokyo 204-8555, Japan}
\address[Showa]{Showa Pharmaceutical University, Machida, Tokyo 194-8543, Japan}
\address[RCNP]{Research Center for Nuclear Physics (RCNP), Osaka University, Ibaraki, Osaka, 567-0047, Japan}
\address[RIKEN]{Theoretical Research Division, Nishina Center, RIKEN, Hirosawa, Wako, Saitama 351-0198, Japan}
\address[JPARC]{J-PARC Branch, KEK Theory Center, Institute of Particle and Nuclear Studies, KEK, Tokai, Ibaraki, 319-1106, Japan
}
\begin{keyword}
{hidden-charm pentaquark; color-octet baryon; exotic hadron; multiquark hadron; baryon-meson scattering}
\end{keyword}
\date{\today}
\begin{abstract}
The $I(J^P)=\half1(\half1^-)$, $\half1(\half3^-)$, and $\half1(\half5^-)$ $uudc\cbar$ pentaquarks 
are investigated by the quark cluster model.
This model, which reproduces
the mass spectra of the color-singlet $S$-wave $q^3$ baryons and $q\qbar$ mesons,
also enables us to evaluate the quark interaction in the color-octet $uud$ configurations.
It is shown that the color-octet isospin-\half1 spin-\half3 $uud$ configuration 
gains an attraction.
The $uudc\cbar$ states with this configuration 
have structures around the $\Sigc{}^{(*)}\Dbar{}^{(*)}$
thresholds:
one bound state, two resonances, and one large cusp are found. 
We argue that the negative parity pentaquark found by the LHCb experiments 
may be given by these structures.
\end{abstract}
%\maketitle
\end{frontmatter}

\section{Introduction}

In 2015, LHCb collaboration  reported that 
two candidates of the new exotic baryons, $P_c$(4380) and $P_c$(4450), 
had been observed in the $\Lambda_b^0\rightarrow \Jpsi p K^-$ decay. 
The mass of the higher peak, $P_c$(4450), 
is 4449.8$\pm$1.7$\pm$2.5 MeV with a width of 39$\pm$5$\pm$19 MeV,
while the lower and broader peak, $P_c$(4380), 
has a mass of 4380$\pm$8$\pm$29 MeV and a width of 205$\pm$18$\pm$86 MeV.
The most favorable set of the spin parity for the lower and the higher peaks is  
$J^P = (\half3^-,\half5^+)$,
though $(\half3^+,\half5^-)$ or $(\half5^+,\half3^-)$ are also acceptable \cite{Aaij:2015tga}.
Also, because the $\Jpsi p$ contribution is necessary to describe the decay data \cite{Aaij:2016phn}, 
it is almost certain
the peaks have the $c\cbar$ pair and are considered 
as the isospin-$\half1$  $uudc\cbar$ pentaquarks.

Beside of the predicting work \cite{Wu:2010jy},
the LHCb observation of $P_c$(4380) and $P_c$(4450) 
has evoked many theoretical studies: the hadronic molecule 
with the meson exchange interaction, the chiral unitary approach with the hidden local gauge symmetry,
the QCD sum rule, the chiral quark model, 
the diquark/triquark model as well as that of the kinematical effects \cite{Chen:2016qju}.
At present, the theoretical and experimental knowledge
is
not enough and one cannot draw a definite picture of these peaks. 

Here, we concentrate our attention on the short range part of 
the hidden-charm pentaquark structure, which is governed by the quark and gluon dynamics. 
For this purpose, we employ the quark cluster model, 
which successfully explained the short range part of the baryon-baryon interaction 
\cite{Oka:2000wj} and
the structure of the light flavored pentaquark $\Lambda(1405)$  \cite{Takeuchi:2007tv}.
It is also shown that
the baryon-baryon interaction derived from the lattice QCD
is found to be similar to that of the quark cluster model 
 \cite{Sasaki:2015ab}. 
Since we are interested in the short range region,
we have investigated the $S$-wave five-quark systems 
as a first step.
They correspond to the negative-parity pentaquarks.
In order to discuss the positive-parity pentaquark state,
which has also been observed by the LHCb experiments,
one has to investigate the $P$-wave five-quark systems,
which is beyond the scope of the present paper.
%Other theoretical works:
%quark model \cite{Huang:2015uda}
%chiral quark model \cite{Yang:2015bmv}
%diquarks \cite{Anisovich:2015zqa}
%quark model review\cite{Chen:2016qju}
%
%meson-exchange \cite{Chen:2015loa}
%hadron model? \cite{Roca:2015dva,Roca:2016tdh}\cite{He:2015cea}\cite{Kahana:2015tkb}
%$\chi_{c1}\, p$ parity negative \cite{Meissner:2015mza}
%soliton\cite{Scoccola:2015nia}
%QCDSR \cite{Wang:2015ava}
%heavy quark limit \cite{Kopeliovich:2015vqa}
%
%triangle singularity \cite{Mikhasenko:2015vca}
%\\
%1\\
%2\\
%3\\
%4\\
%5\\
%6\\
%7\\
%8\\
%9\\
%10

\begin{table}[t]
\renewcommand\arraystretch{1.4}
\tabcolsep=1mm
\caption{The classification of the isospin-$\half1$ negative parity $qqqc\cbar$ states.
The $uud$ spin ($s_q$), color ($c$), 
CMI of the five quark systems at the heavy quark limit 
($\bra \Ocmi \ket_{5q}^{(HQ)}$), the $c\cbar$ spin ($s_c$), the total spin of  $uudc\cbar$ ($J$),
the lowest $S$-wave threshold (T) and the CMI contribution to the threshold energy ($\bra \Ocmi \ket_\text{T}^{(HQ)}$) are listed.}
\begin{center}
\begin{tabular}{ccccccccccccccccccccccc}\hline
& $s_q$  & $c$  & $\bra \Ocmi \ket_{5q}^{(HQ)}$  &  $s_c$ & $J$ & T & $\bra \Ocmi \ket_\text{T}^{(HQ)}$
\\\hline
\cwsh\  &$\half1$ & {\bf 1} &$-8$& 0 & $\half1$ &~~~$N\etac$~~~ & $-8$
\\
&& && 1 & $\half1$ &$N\Jpsi$&$-8$
\\
&& && 1 & $\half3$ &$N\Jpsi$&$-8$
\\\hline
\crsh\ & $\half1$ & {\bf 8} &$-2$& 0 & $\half1$ & \Lamc\Dbar&$-8$
\\
&& && 1 & $\half1$ & \Lamc\Dbar&$-8$
\\
& &&& 1 & $\half3$ &  \Lamc\Dbarstar& ${-8}$
\\\hline
\crsth\ & $\half3$ & {\bf 8} &\phantom{$-$}2& 0 & $\half3$ &   \Sigcstar\Dbar& ${8\over 3}$
\\
&& && 1 & $\half1$  &   \Sigc\Dbar&
${8\over 3}$
\\
&& && 1 & $\half3$ &  \Sigcstar\Dbar& ${8\over 3}$
\\
& &&& 1 & $\half5$ &   \Sigcstar\Dbarstar& ${8\over 3}$
\\\hline
\end{tabular}
\end{center}
\label{tbl:hq}
\end{table}%

Let us first discuss possible configurations of $uud$ quarks in the $uudc\cbar$ systems.
Since the whole system is the color-singlet  and 
%since 
the $c\cbar$ pair
is color-singlet or octet,
 the remaining three light quarks are also color-singlet or color-octet.
So, when the orbital configuration is totally symmetric,
the $uud$ configuration in the $uudc\cbar$ systems is totally symmetric ({\bf 56}-plet)
or mixed symmetric ({\bf 70}-plet) in the flavor-spin
SU$_{f\sigma}$(6) 
space. They are classified as:
\begin{align}
{\bf 56}_{f\sigma} 
&=
  {\bf 8}_f \times {\bf 2}_\sigma
+ {\bf 10}_f\times {\bf 4}_\sigma
\\
{\bf 70}_{f\sigma} 
&= 
  {\bf 1}_f \times {\bf 2}_\sigma
+ {\bf 8}_f \times {\bf 2}_\sigma
+ {\bf 8}_f \times {\bf 4}_\sigma
+ {\bf 10}_f\times {\bf 2}_\sigma\ .
\end{align}
Here the numbers are the dimension of the corresponding representations.
The color-singlet $uud$ systems
correspond to the usual {\bf 56}-plet baryons. 
The color-octet {\bf 70}-plet systems
can be decomposed into 
the flavor-singlet spin-$\half1$ (${\bf 1}_f\times {\bf 2}_\sigma$), 
the flavor-octet spin-$\half1$ (${\bf 8}_f\times {\bf 2}_\sigma$), 
the flavor-octet spin-$\half3$  states (${\bf 8}_f\times {\bf 4}_\sigma$),
and the flavor-decuplet spin-$\half1$ (${\bf 10}_f\times {\bf 2}_\sigma$).
Since the present work concerns systems of the isospin $\half1$ and the strangeness zero,
namely, flavor-octet systems,
the configurations of the three light quarks correspond to  
one of the following three:
(a)  color-singlet spin-$\half1$, 
(b)  color-octet spin-$\half1$,  and
(c)  color-octet spin-$\half3$,
each of them we denote by \cwsh, \crsh, and \crsth\ in the following, respectively.
Since the spin of the $c\cbar$ pair is either 0 or 1, 
the total spin of the $uudc\cbar$ systems is
either $\half1$ (5-fold), $\half3$ (4-fold), or $\half5$ (1-fold). (See Table \ref{tbl:hq}.)

In the short range part of the two-hadron interaction, the color-magnetic interaction (CMI) 
plays an important role.
We evaluate the color flavor spin part of CMI,
\begin{align}
\Ocmi&=-\sum_{ij}{m_u^2\over m_im_j}\lamilamj\sigisigj \ ,
\label{eq:Ocmi}
\end{align}
by the quark wave function.
In eq.\ (\ref{eq:Ocmi}),
$m_i$, $\lambda_i$, and $\sigma_i$ are
the constituent mass, the Gell-Mann matrix in the color space,
and the Pauli spin matrix for the $i$th (anti)quark, respectively.
Because of the factor ${m_u^2\over m_im_j}$, only the operators between the light quarks give non-zero
contribution at the
heavy quark limit. 
So, CMI estimated 
by the above three-light-quark configurations
 actually correspond to the estimates of the whole $uudc\cbar$ 
at that limit, $\bra \Ocmi\ket_{5q}^{(HQ)}$,
which is listed in Table \ref{tbl:hq}.
There,  we also show the lowest $S$-wave baryon-meson threshold for each state
together with the contribution of   
$\Ocmi$ to that threshold at the heavy quark limit:
\begin{align}
\bra \Ocmi\ket_\text{T}^{(HQ)}&=\bra \Ocmi\ket_B^{(HQ)}+\bra \Ocmi\ket_M^{(HQ)}\ ,
\end{align}
where $\bra \Ocmi\ket_B^{(HQ)}$ and $\bra \Ocmi\ket_M^{(HQ)}$ are $\Ocmi$ evaluated 
by the baryon and the meson wave functions
at the heavy quark limit, respectively.
Suppose we estimate the baryon-meson potential arising from CMI by
\begin{align}
V_\text{cmi}^{\text{eff}(HQ)} &\sim
\bra \Ocmi\ket_{5q}^{(HQ)} -\bra \Ocmi\ket_\text{T}^{(HQ)},
\label{eq:Veff}
\end{align}
then only those which include the \crsth\ configuration
is attractive
though its energy is actually the highest.
As in Table \ref{tbl:hq},
there are four states which gain such an attraction
in the isospin-\half1 $uudc\cbar$ systems.
Since $uudc\cbar$ is color-singlet as a whole, 
the system of the color-octet $uud$ with color-octet $c\cbar$ 
can be observed as \Lamc\Dbar${}^{(*)}$ or \Sigc${}^{(*)}$\Dbar${}^{(*)}$ baryon meson states,
where each of the hadrons is color-singlet.
The attraction within the color-octet spin-\half3 $uud$ is observed
as the attraction between the $\Sigma_c^{(*)}$ baryon and the $\Dbar^{(*)}$ meson.
In this sense, when this interaction causes pentaquark states, 
one may call them `hidden color-octet $uud$ baryons.'

As we will show in this letter, 
the above behavior remains visible even after we employ the realistic quark masses and perform 
dynamical calculations.
We have found that there is a bound state
in the $J=\half5$ system, 
a resonance and a cusp in the $\half1$ system, and a resonance in $\half3$ system;
which exactly correspond to those of the \crsth\ configuration
as listed in Table \ref{tbl:hq}.
Since the expectation values of CMI by the $uud$ configurations
can be calibrated by the observed hadron masses,
and since they do not depend on the heavy quark mass,
the above four structures are robust to change of the parameters.
Though we use a rather complicated model Hamiltonian in the following
in order to produce the threshold energies correctly,
the results do not depend much on the model details;
the situation is the same, for example, 
when the system goes to the bottom sector.
Thus, we would like to argue that the negative parity peak of the LHCb pentaquark
% $P_c(4380)$ or $P_c(4450)$ 
 may consist of (some of) these structures caused by the color-octet $uud$ configuration. 

\section{Method}

%\subsection{Model hamiltonian}

We employ the coupled-channel quark cluster model to investigate the
$uudc\cbar$ $I(J^P)=\half1(\half1^-,\half3^-,\half5^-)$ systems.
This model becomes a $(0s)^5$ quark model in the baryon-meson short range region.
In the long range region,
this model becomes 
essentially 
a baryon-meson model.
There, the interaction between the baryon and the meson
arises from the quark degrees of freedom and from the interaction between quarks.

The model Hamiltonian, 
$H_q$, consists of the central spin-independent term, $H_c$,
and the color spin term, $\Vcmi$.
The $H_c$ consists of 
the kinetic term, $K$, the confinement term, $\Vconf$, and the 
color Coulomb term, $\Vcoul$:
\begin{align}
H_q&=H_c+\Vcmi 
\\
H_c&= K+\Vconf+\Vcoul\ .
\end{align}
Both of $\Vcoul$ and $\Vcmi$ come from the effective one-gluon exchange interaction
between quarks.
Each of the terms is taken to be slightly different from the
conventional quark model \cite{Godfrey:1985xj,Capstick:1986bm}.
It is because we use a single Gaussian 
for the orbital wave function of each of the $q\qbar$ and $q^3$ hadrons
in order to make it feasible to solve the five-quark systems.
%\subsubsection{Kinetic term}

The kinetic term is taken as nonrelativistic:
\begin{align}
K   &=\sum  m_i+{1\over 2m_i}\Big(\vecp_i-{m_i\over M_G}\vecP_G\Big)^2\ ,
\end{align}
where $\vecp_i$ is the momentum of the $i$th (anti)quark, and
$M_G$ and $\vecP_G$ are the total mass and momentum of the five-quark system.

%\subsubsection{Confinement term}

The linear confinement term is 
\begin{align}
\Vconf&=\sum_{i<j} \lamilamj\,(  -a_c r_{ij} + c_1+{c_2^2\over \mu_{ij}}+c_{q\qbar} )\ .
\end{align}
The value of the confinement strength, $a_c$, is taken from the
Lattice QCD calculation \cite{Kawanai:2011xb},
whose value corresponds 1.12 GeV$^2$ for the $q\qbar$ systems.
The $r_{ij}$ and $\mu_{ij}$ are the relative distance and the reduced masses of the $i$th and the $j$th quarks, respectively.
In the above equation, the $c_1$, $c_2$, and $c_{q\qbar}$
are the constant parameters.
The $c_1$ and $c_2$ express the constant mass shift
which is expanded up to the $\mu_{ij}^{-1}$ term.
The parameter $c_{q\qbar}$ is nonzero only when this operates on quark-antiquark pairs.
We use these constants as  
free parameters so that the model produces the observed masses of the 
relevant single hadrons, which have a quite wide energy range.
The constant mass shift itself appears
when the potential is constructed from the lattice QCD calculation,
though the values are different \cite{Kawanai:2011xb}.

%\subsubsection{Color-Coulomb term}

The color-Coulomb term is written as
\begin{align}
\Vcoul&=\sum_{i<j}{\lamilamj\over 4}{\alpha_s(r_{ij})\over r_{ij}}
\\
%\alpha_s(Q) &= \sum_{k=1}^3 \alpha_k \exp[-{Q^2\over 4\gamma_k^2}]  \ .
\alpha_s(r) &= \sum_{k=1}^3 \alpha_k \,\text{erf}[\gamma_k r]  \ .
\end{align}
The strong coupling constant, $\alpha_s$, 
is assumed to depend on the relative distance of the interacting quarks,
following the manner of refs.\ \cite{Godfrey:1985xj,Capstick:1986bm}.
In this work, $\alpha_s$ at small $Q^2$
is refitted so that it corresponds to the running coupling constant for $Q^2>3$ GeV
in the momentum space, 
whereas it goes to 0.8 at $Q^2=0$:
($4\gamma_k^2$ in GeV$^2$, \ $\alpha_k$)=(1.5, 0.45), (10, 0.15), (1000, 0.20).

%\subsubsection{Color magnetic term}

The CMI term is
\begin{align}
\Vcmi &= -\sum_{i<j}{\lamilamj\over 4}\alpha_s^{ss}{}_{ij}{2\pi\over 3m_im_j}\sigisigj\delta^3(\vecr_{ij})
\\
\alpha_s^{ss}{}_{ij}&=\begin{cases}\displaystyle  \alpha_{s1}^{ss}+ \alpha_{s2}^{ss}{m_u\over \mu_{ij}}& \text{for a $qq$ pair}\\
\alpha_{s3}^{ss}& \text{for a $q\qbar$ pair.}
\end{cases}
\end{align}
We use $\alpha_{s1}^{ss}$ and $\alpha_{s2}^{ss}$ as parameters to 
fit the contribution of the $uu$ pairs in the baryons such as
$2m_{\Sigma_Q^*}+m_{\Sigma_Q}-3m_{\Lambda_Q}$ and that of the $uQ$ pairs such as
$m_{\Sigma_Q^*}-m_{\Sigma_Q}$.
The value of $\alpha_{s3}^{ss}$ is determined by taking an average of the \Dbarstar-\Dbar\ and $D_s^*$-$D_s$ mass differences.
Again these mass dependent coupling constants have to be introduced 
so that the model gives the correct hyperfine splitting of the hadron masses.
In this way, we calibrate the size of the CMI,
which is the origin of the attraction focused in this work,
from the observables.
The parameters are summarized in Table \ref{tbl:param}.

\begin{table}
\caption{Model parameters. The quark masses, $m_Q$, and the constant parameters, $c_i$, are 
in MeV.}
\begin{center}
\renewcommand\arraystretch{1.2}
\begin{tabular}{cccccccccccccc}\hline
$m_u(=m_d)$ &$m_c$ &$c_1$ &$c_2$&$c_{q\qbar}$\\  \hline
300 &1741.5&86.4& 113.9 &$-$5.65\\ \hline
$a_c$(MeV/fm) 
&$\alpha_{s1}^{ss}$&$\alpha_{s2}^{ss}$&$\alpha_{s3}^{ss}$ \\  \hline
196.9 &$-$1.0967 & 0.4756 &0.5668\\ \hline
\end{tabular}
\end{center}
\label{tbl:param}
\end{table}%

%\subsection{Wave functions}

%\subsubsection{The spin-flavor-color part}

The color flavor spin part of the $q^3$ or $q\qbar$ wave functions is
taken as a conventional way  \cite{Stancu:1996,pdg}.
The orbital wave function of the mesons, $\phi_M$, and that of the baryons, $\phi_B$,
are written by Gaussian with a size parameter $b$, $\phi(\vecr,b)$:
\begin{align}
\phi_M(\vecr_M) &= \phi(\vecr_{12},{x_{0}\over \sqrt{\mu_{12}}})
\\
\phi_B(\vecr_B) 
&=\phi(\vecr_{12},{x_{0}\over \sqrt{\mu_{12}}}) \phi(\vecr_{12-3},{x_{0}\over \sqrt{\mu_{12-3}}})
\ ,
\end{align}
where the reduced masses, $\mu_{12}$ and $\mu_{12-3}$, 
correspond to the Jacobi coordinates, $\vecr_{12}$ and $\vecr_{12-3}$.
We assume that the size parameter of the orbital motion can be 
approximated by $b = x_0/\sqrt{m}$ and
minimize the central part of the Hamiltonian, $H_c$, against 
$x_0$ for each flavor set: 
$u\cbar$, $c\cbar$, $uud$, $udc$.
For the baryons, this means that the ratio of the size parameters is kept to a certain mass ratio;
 {\it e.g.}, $b_{uc}/b_{ud}$
in \Lamc\ or \Sigc\ is equal to $\sqrt{\mu_{ud}/\mu_{uc}}$. %$\sqrt{(m_u+m_c)/2m_c}$.

\begin{table}
\caption{Single hadrons masses obtained by the present model. The isospin-averaged masses are taken from \cite{pdg}.}
\begin{center}
\renewcommand\arraystretch{1.2}
\begin{tabular}{ccccc}\hline
Baryon         &
N              &
$\Lambda_c$    &
$\Sigma_c$     &
$\Sigma_c^*$   \\\hline
Mass&    
~922.3& 
2291.8& 
2453.6& 
2516.0\\
Obs.\ &
~938.9& 
2286.5& 
2453.5& 
2518.1\\ \hline 
Meson &
$\eta_c$       &
\Jpsi\         &
\Dbar\         &
\Dbarstar\     \\\hline
Mass&    
2981.3&
3100.9&
1863.4&
2004.9\\
Obs.\ &
2983.6&
3096.9&
1867.2&
2008.6\\
\hline
\end{tabular}
\label{tbl:cmibm}
\end{center}
\end{table}

%\subsubsection{The masses and size of the hadrons}

The masses of the relevant hadrons obtained by the models are summarized in table \ref{tbl:cmibm}
together with the observed masses.
The obtained $x_0$'s are listed in Table \ref{tbl:x0}
with the corresponding size parameters.
It is found that  $x_0$ does not vary much while
the difference between $b_{uu}$, $b_{uc}$ and $b_{cc}$ is large.
In order to investigate systems with more than one charm quark
one needs to take into account the flavor dependence of the orbital motion.

\begin{table}
\caption{The size parameter $b_{ij}$ (fm) and the parameter $x_0$ (fm$^{1/2}$) obtained by minimizing the central part of the Hamiltonian, $H_c$, for each of the systems. }
\begin{center}
\begin{tabular}{lcccccccccccccc}\hline
system   & $x_0$&$b_{uu}$&$b_{uc}$&$b_{cc}$\\\hline
$uud$    & 0.60 & 0.68   &        &        \\
$uuc$    & 0.62 & 0.71   & 0.54   &        \\
$ucc$    & 0.65 &        & 0.57   & 0.31   \\
$u\cbar$ & 0.56 &        & 0.49   &        \\ 
$c\cbar$ & 0.61 &        &        & 0.29   \\ \hline
\end{tabular}
\end{center}
\label{tbl:x0}
\end{table}%

%\ref{tbl:singlehadrons}

%\subsubsection{Baryon-meson system}

We employ the resonating group method (RGM) to solve the five-quark systems.
The wave function, $\Psi$, 
can be expanded by the locally peaked Gaussians for each of
the baryon-meson channel $\nu$ as \cite{Oka:2000wj,Takeuchi:2007tv}
\begin{align}
\Psi &= \sum_{\nu,i} c^\nu_i
{\cal A}_q \big\{\psi_B^\nu(\vecr_B)\psi_M^\nu(\vecr_M)\chi(\vecR,\vecSp_i)\big\}
\label{eq:16}
\\
\chi(\vecR,\vecSp_i)&=i_0({1\over b^2}\vecR\cdot\vecSp_i)\exp[-{1\over 2b^2}(R^2+S_i^2)]
\ ,
\label{eq:17}
\end{align}
where ${\cal A}_q$ stands for the quark antisymmetrization, which operates on the four quarks, 
and $i_\ell(z)$ is the modified spherical Bessel function.
As for the scattering state, the wave function of the relative motion 
is connected smoothly to the spherical Hankel functions in the long range region.

By integrating out the internal wave function of the hadrons, 
the RGM equation can be obtained from the equation of motion for the quarks 
$(H_q-E) \Psi  = 0$, as
\begin{align}
\sum_{\nu'j} (H_{ij}^{\nu\nu'}-EN_{ij}^{\nu\nu'}) c^{\nu'}_j=0
\end{align}
with the Hamiltonian and normalization kernels
\begin{align}
\left\{
\begin{matrix}
H_{ij}^{\nu\nu'}\\N_{ij}^{\nu\nu'}
\end{matrix}
\right\}
&=
\int\rmd \vecr \rmd\vecr'\;
  \psi_B^{\nu\dag}(\vecr_B)\psi_M^{\nu\dag}(\vecr_M)\chi^\dag(\vecR,\vecSp_i)
\nonumber\\
&\times
\left\{
\begin{matrix}
 H_q \\1
\end{matrix}
\right\}
 {\cal A}_q 
\big\{\psi_B^{\nu'}(\vecr'_B)\psi_M^{\nu'}(\vecr'_M)\chi(\vecR',\vecSp_j)\big\} 
\ ,
\label{eq:RGMk}
\end{align}
where d$\vecr$ stands for the integration over all the Jacobi coordinates of the five-quark system.
For the detail of the calculation, see, for example, appendix B of ref.\ \cite{Takeuchi:2007tv}.
We choose the parameters $b$ and $\vecSp_i$ in eqs.\ (\ref{eq:16}) and
(\ref{eq:17})
so that the calculating results are stable against changing the parameters.
Thus, after fitting single hadron masses, the model is essentially parameter-free.

Here we define a three-body operator, $\calPcs$, 
to extract the $uud$ color-$c$, spin-$s$, orbital $(0s)^3$ configuration
from the resonance as well as from the bound states
in order to evaluate its size.
It is defined as
\begin{align}
\calPcs{}_{cs}  &= |uud;cs(0s)^3\ket \bra uud;cs(0s)^3|
\\
\calPcs&=\sum_{c,s}\calPcs_{cs}
\ .
\end{align}
We use the same value as that of $b_{uu}$ in $\Sigma_c$ for the size parameter of the $(0s)^3$ component.

%%%%%%%%%%%%%%

\section{Results and Discussions}

\subsection{$uudc\cbar$ $I(J^P)$=$\half1(\half5^-)$ state}

Suppose the orbital part of the $uudc\cbar$ system is
totally symmetric, such as the system in the orbital $(0s)^5$ configuration,
the color-spin-flavor part of the $uudc$ quarks should be totally antisymmetric.
For the $S$-wave $uudc\cbar$ $I(J^P)$=$\half1(\half5^-)$ channel,
this state is  
\Sigcstar\Dbarstar\
antisymmetrized over the quarks.
The color flavor spin part of its
normalization, $\bra \Sigcstar\Dbarstar |\calA_q|\Sigcstar\Dbarstar\ket$, is ${4\over 3}$.
For a channel which has a repulsion from Pauli blocking over the quarks,
this normalization becomes smaller than 1.
When the normalization is larger than 1, like this case,
an attraction between the two hadrons arises in the short range region.
Moreover, the $c\cbar$ pair in this state is spin-1 and the three light quarks are 
color-octet spin-\half3,
\crsth.
The attraction is expected to come also from CMI. 

%\subsubsection{Baryon-meson dynamical model}

We investigate the \Sigcstar\Dbarstar\ $\half1(\half5^-)$ channel 
by the quark cluster model.
A very shallow bound state is found
1.0 MeV below the threshold.
The probability to find the  $(0s)^3$ $uud$ 
in this bound state, $\bra \calP\ket$, 
is found to be 0.21.
In this channel,  
 the $uud$ quarks in the $(0s)^3$ configuration are all color-octet spin-\half3.
Suppose there is no
 Pauli exclusion principle applied between the $ud$ quarks in \Sigc\ and the $u$ quark in \Dbarstar,
the ratio of the color-singlet to the color-octet probabilities should be 1 to 8.
The color-singlet configuration vanishes by introducing the quark antisymmetrization,
 and this small change of the configuration in size, $\sim 0.21\times$1/9, 
induces the bound state.

\subsection{$uudc\cbar$ $I(J^P)$=$\half1(\half3^-)$ states}

The $uudc\cbar\ \half1(\half3^-)$ states consist of five baryon-meson channels:
$N\Jpsi$, \Lamc\Dbarstar, \Sigcstar\Dbar, \Sigc\Dbarstar, and \Sigcstar\Dbarstar.
When the baryon and the meson come close to the overlapping region,
these channels are not orthogonal to each other any more.
There are 
four $uud$ $\half1(\half3^-)$
 states, \cwsh, \crsh, and \crsth\ (2-fold),
whereas five baryon-meson channels exist.
Thus one forbidden baryon-meson state, whose norm is zero, appears
when the system is 
totally symmetric in the orbital space.
All the diagonal elements of the color flavor spin part of the normalization, however,
 are close to or larger than 1;
the baryon-meson single channels do not gain 
the large repulsion from the quark Pauli-blocking.

We have performed the quark cluster model calculation 
where all the relevant five baryon-meson channels are coupled
for the $S$-wave $uudc\cbar$ $\half1(\half3^-)$ channel.
Before that, however,
we first discuss the results of the three-channel calculation of the \Sigcstar\Dbar,
\Sigc\Dbarstar,  and \Sigcstar\Dbarstar\ 
in order to see the effects of the \crsth\ configuration.
It is found that 
a bound state  appears 0.08 MeV below the \Sigcstar\Dbar\ threshold.
Also, a very sharp resonance with the width less than 0.1 MeV 
is found in the \Sigcstar\Dbar\ channel at the energy 
0.8 MeV below the \Sigc\Dbarstar\ threshold.
The probability of the $uud$ to form one of the $(0s)^3$ configurations, $\bra\calPcs\ket$,
in the bound state 
is found to be 0.08.
The proportion of the factor to find each \cwsh, \crsh, and \crsth\ configurations,
$\bra \calPcs_{cs}\ket/\bra\calPcs\ket$
%{}_{1{1\over 2}}\ket +\bra \calPcs{}_{8{1\over 2}}\ket +\bra \calPcs{}_{8{3\over 2}}\ket $ 
is
listed in Table \ref{tbl:prob0s} with an identification of C$'$,
together with those of the resonance (B$'$).
Those proportions are very similar to the
antisymmetrized \Sigcstar\Dbar\ (0.05 0.19 0.76) and \Sigc\Dbarstar (0.01 0.06 0.93).
Thus, the bound state C$'$ is considered 
essentially as the antisymmetrized \Sigcstar\Dbar, and
the resonance B$'$
is considered as the antisymmetrized
\Sigc\Dbarstar.

\begin{table}[tb]
\caption{The bound state, resonances and cusp obtained by the present model.
The four structures are identified by the letter A-D in the text. The identification with a dash (B$'$-D$'$) is used for the result of the \Sigc${}^{(*)}$\Dbar${}^{(*)}$ three-channel calculation. 
The energies, $E$, are shown in MeV. 
The proportions of the factors to find each of \cwsh, \crsh, and \crsth,
$\bra \calPcs_{cs}\ket/\bra\calPcs\ket$, are listed under the entry [$q^3cs$].
}
\begin{center}
\tabcolsep=0.9mm
\renewcommand\arraystretch{1.2}
\begin{tabular}{lllcrrrrrr} \hline
id.&\multicolumn{2}{l}{initial channel $(J^P)$}& $E$ & \cwsh & \crsh & \crsth  \\\hline
A&\Sigcstar\Dbarstar   $(\half5^-)$ & bound state  &  4519.9 & 0.00&0.00&1.00\\
B&$N$\Jpsi\            $(\half3^-)$ & cusp         &  4458.0 & 0.21&0.02&0.77\\ 
C&$N$\Jpsi\            $(\half3^-)$ & resonance    &  4379.3 & 0.24&0.16&0.60\\ 
D&$N$\Jpsi\            $(\half1^-)$ & resonance    &  4316.5 & 0.75&0.13&0.12\\\hline 
B$'$&\Sigcstar\Dbar\   $(\half3^-)$ & resonance    &  4457.8 & 0.02&0.08&0.90\\
C$'$&\Sigcstar\Dbar\   $(\half3^-)$ & bound state  &  4379.3 & 0.05&0.21&0.74\\
D$'$&\Sigc\Dbar\       $(\half1^-)$ & bound state  &  4317.0 & 0.05&0.22&0.73\\\hline
\end{tabular}
\end{center}
\label{tbl:prob0s}
\end{table}%

\begin{figure}[btp]
\begin{center}
\includegraphics[width=9cm]{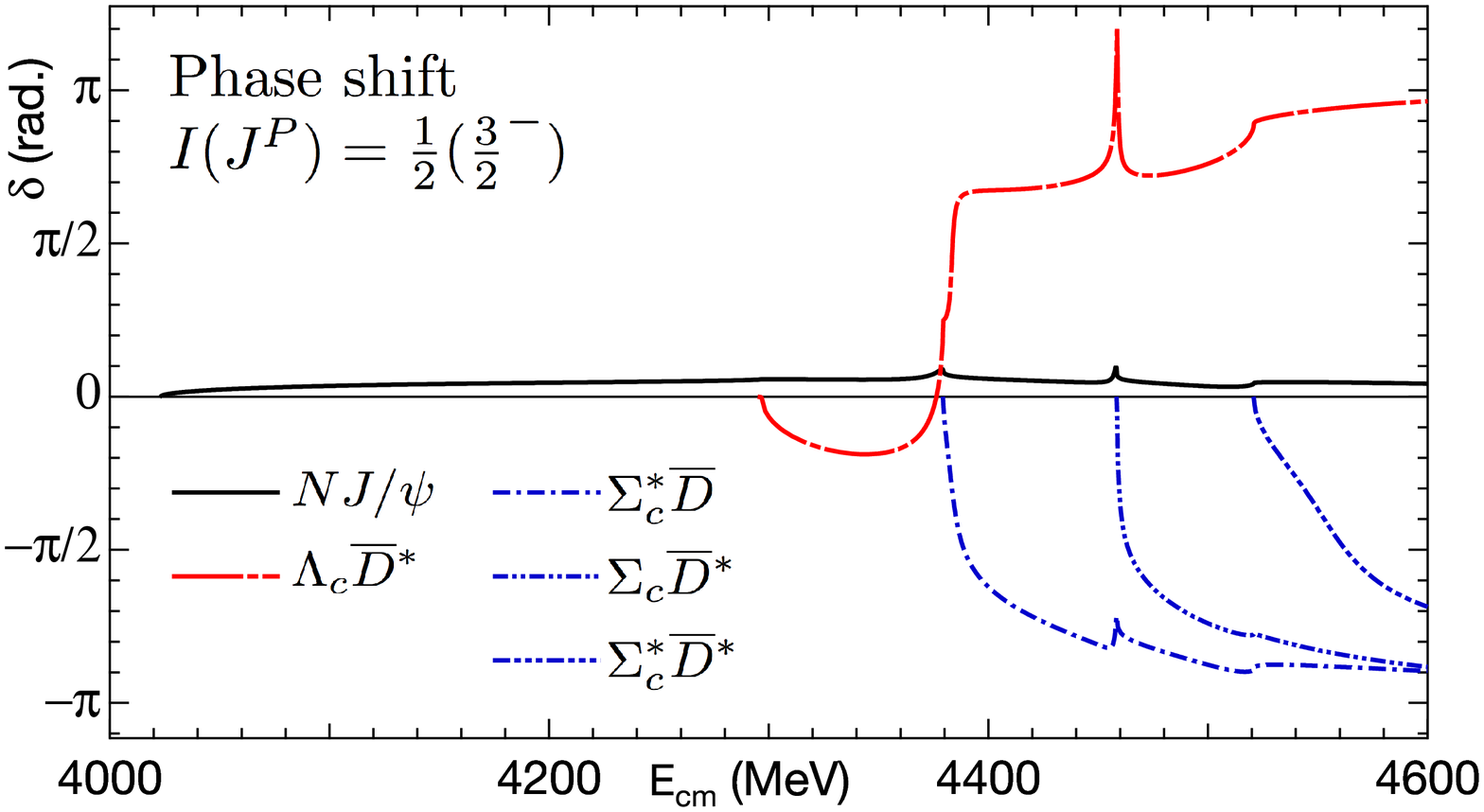}
\caption{The diagonal scattering phase shifts of the $S$-wave $uudc\cbar$ $I(J^P)$=$\half1(\half3^-)$ system.
The solid line is that of the $N\Jpsi$ channel, 
the long-dot-dashed line is for the \Lamc\Dbarstar channel,
and the dot-dashed, double-dot-dashed, and  triple-dot-dashed lines are 
for the \Sigcstar\Dbar, \Sigc\Dbarstar, and \Sigcstar\Dbarstar channels,
respectively. (Color online.)}
\label{fig:ps23}
\end{center}
\end{figure}

In figure \ref{fig:ps23},
we show phase shifts of the $S$-wave $uudc\cbar$ $\half1(\half3^-)$ 
by the calculation where all five baryon-meson channels are coupled.
The plotted phase shifts are the diagonal ones;
namely, the initial and the final channels are taken to be the same.
%At the resonance energy one of these phase shifts increases up to  above $\pi/2$.
%
The bound state and the resonance found in the above three-channel calculation
now become a sharp resonance and a cusp in the \Lamc\Dbarstar\ channel, respectively.
The energy of the resonance or the cusp does not move much when the $N\Jpsi$ and \Lamc\Dbarstar\ channels are introduced.
Their energies are summarized in Table \ref{tbl:prob0s}.
Among the $\bra \calPcs_{cs}\ket/\bra\calPcs\ket$ evaluated by the scattering wave function with the initial $N$\Jpsi\ channel,
the factor to find \crsth\ is the largest
in both of the resonance and in the cusp.
In table \ref{tbl:prob0s}, we list their proportions at the resonance or the cusp energy
(with identifications of B and C).
In the five-channel calculation, the proportion of \cwsh\ becomes larger than that of the three-channel calculation, but still the \crsth\ component is the largest.
The listed energy of each of the structures is that where the $\bra \calP\ket$ becomes local maximum.
All the resonance and cusp energies read from the phase shifts differ by less than 1 MeV from 
the listed ones except for the resonance C, 
where the phase shift increases up to above $\pi/2$ at by 4 MeV above the listed energy.

\subsection{$uudc\cbar$ $I(J^P)$=$\half1(\half1^-)$ states}

The $uudc\cbar\ \half1(\half1^-)$ states consist of seven baryon-meson channels:
$N\etac$, $N\Jpsi$, \Lamc\Dbar, \Lamc\Dbarstar, \Sigc\Dbar, \Sigc\Dbarstar, 
and \Sigcstar\Dbarstar,
whereas there are five $(0s)^5$ states.
So, there are two forbidden states when the system is 
totally symmetric in the orbital space.
Also in this case, all the diagonal elements
of the normalization are close to 1;
no baryon meson
state is affected strongly by the quark Pauli-blocking.

Again, we first discuss the results of the three-channel quark cluster model calculation:
\Sigc\Dbar,
\Sigc\Dbarstar, and
\Sigcstar\Dbarstar.
There is one bound state with the binding energy 0.13 MeV,
but no resonance is found.
As is seen from the entry with an identification D$'$ in Table \ref{tbl:prob0s}, major component of this bound state is \crsth.
In the antisymmetrized \Sigc\Dbar\ state, the proportion is (0.05 0.19 0.76).
Thus this bound state is essentially an antisymmetrized \Sigc\Dbar.

\begin{figure}[bt]
\begin{center}
\includegraphics[width=9cm]{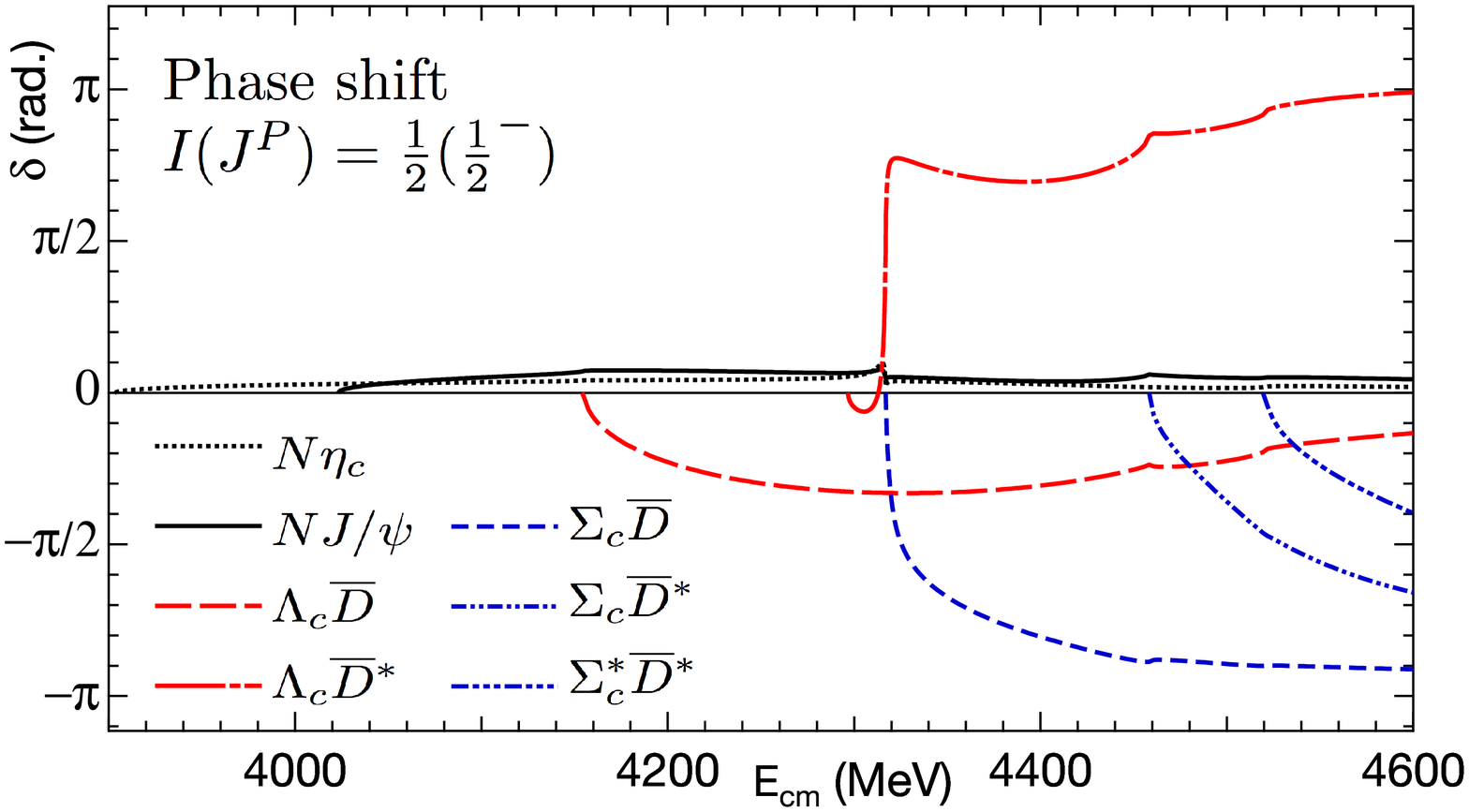}
\caption{The diagonal scattering phase shifts of the $S$-wave $uudc\cbar$ $I(J^P)$=$\half1(\half1^-)$ system.
The solid and the dotted lines are those of the $N\Jpsi$ and the $N\eta_c$ channels, 
the long-dashed and the long-dot-dashed lines are for 
\Lamc\Dbar\ and \Lamc\Dbarstar channels, 
and the dashed,  double-dot-dashed, and  triple-dot-dashed lines are 
for the  \Sigc\Dbar, \Sigc\Dbarstar, and \Sigcstar\Dbarstar channels,
respectively. (Color online.)
}
\label{fig:ps21}
\end{center}
\end{figure}

In figure \ref{fig:ps21},
we show the diagonal phase shifts 
of the 7-channel calculation
of this system.
There is a sharp resonance in the \Lamc\Dbarstar\ channel,
which corresponds to the bound state of the three-channel calculation.
In Table \ref{tbl:prob0s}, we list $\bra \calPcs_{cs}\ket/\bra\calPcs\ket$
at the resonance energy under an identification D.
The factor to find the \crsth\ configuration becomes
small compared to that of the three-channel calculation;
the \cwsh\ configuration becomes the largest at the resonance.
As we will show later,
the coupling to the $N\Jpsi$ channel to the \Lamc\Dbar\ and \Sigc\Dbar\ channels is
stronger in this 
 $\half1(\half1^-)$ case.
The existence of the attraction in the \Sigc\Dbar\ or \crsth, however, is important to create a resonance.
 
%The solid line is that of the $N\Jpsi$ channel, 
%the dotted line the $N\eta_c$, 
%the long-dashed and the long-dot-dashed lines are for 
%\Lamc\Dbar\ and \Lamc\Dbarstar, 
%and the dashed, dot-dashed, double-dot-dashed, and  triple-dot-dashed lines are 
%for the  \Sigc\Dbar, \Sigcstar\Dbar, \Sigc\Dbarstar, and \Sigcstar\Dbarstar,
%respectively. 

%\subsection{Coupling between the resonances and \Jpsi$N$}
\subsection{$uudc\cbar$ pentaquarks}

As summarized in Table \ref{tbl:prob0s},
in the present model,
one bound state is found in $\half1(\half5^-)$, %4520 MeV,
one resonance and a cusp in $\half1(\half3^-)$, and
one resonance in $\half1(\half1^-)$.
As seen in Table \ref{tbl:hq},
these structures exactly correspond to the
number of \crsth\ configurations in the negative parity baryon meson channels.
The energy of the color-octet $uud$ configuration itself 
is higher than that of the color-singlet $uud$ configurations.
CMI works as an attractive force in the \crsth\ configuration
just because its energy is lower than the that of the 
relevant baryon-meson threshold, $\Sigma^{(*)}\Dbar{}^{(*)}$.
So, when the model space is enlarged from the $\Sigma^{(*)}\Dbar{}^{(*)}$ 
to all the relevant baryon-meson systems, for example, D$'$ to D,
the proportion of the \crsth\ configuration becomes smaller,
from 0.73 to 0.12.
The resonance, however, is still there.
%Moreover,
%the proportion of the \crsth\ configuration is found to enhance at around the resonance energy: 0.07 at 4312 MeV, 0.09 at 4317 MeV, and 0.03 at 4322 MeV.
Situation is similar for the resonances B and C,
though the reduction of the proportion
by introducing the lower channels is less extreme.
The \crsth\ configuration plays an important role even though the
proportion becomes small. 
These structures are considered to be an appearance of the hidden color-octet $uud$ baryon.

These resonances and cusp in $(\half1^-,\half3^-)$ exist in the energy range  of the $P_c(4380)$ peak,
$E\pm \Gamma/2 =$  4278--4483 MeV.
All the structures are very close to the baryon meson thresholds,
and each of the resonances has a width of a few MeV,
which is far smaller than the widths of the observed peaks.
%It is because the coupling between the \Lamc\Dbarstar\ channel, where the resonances and the cusp exist,
%and the \Sigc${}^{(*)}$\Dbar${}^{(*)}$ channels, where the \crsth\ configuration exists,
%is small in the present model.
Including the light meson exchange effects 
between the $Y_c$ baryon and the \Dbar\ meson may enlarge the width, 
which is an interesting topic and will be investigated in future works.
We argue that these resonances and cusp may combine to form the broad peak of $P_c(4380)$.
Or, if the parity of $P_c(4450)$ is found to be negative in future experiments,
the cusp at 4458 MeV may correspond to that peak.
The bound state in $\half5^-$, whose energy is higher than both of the observed pentaquark peaks, does not couple to the $S$-wave $N\Jpsi$ channel;
the higher partial wave mode is necessary to 
see this bound state from the $N\Jpsi$ channel.

\begin{figure}[t]
\begin{center}
\includegraphics[width=8cm]{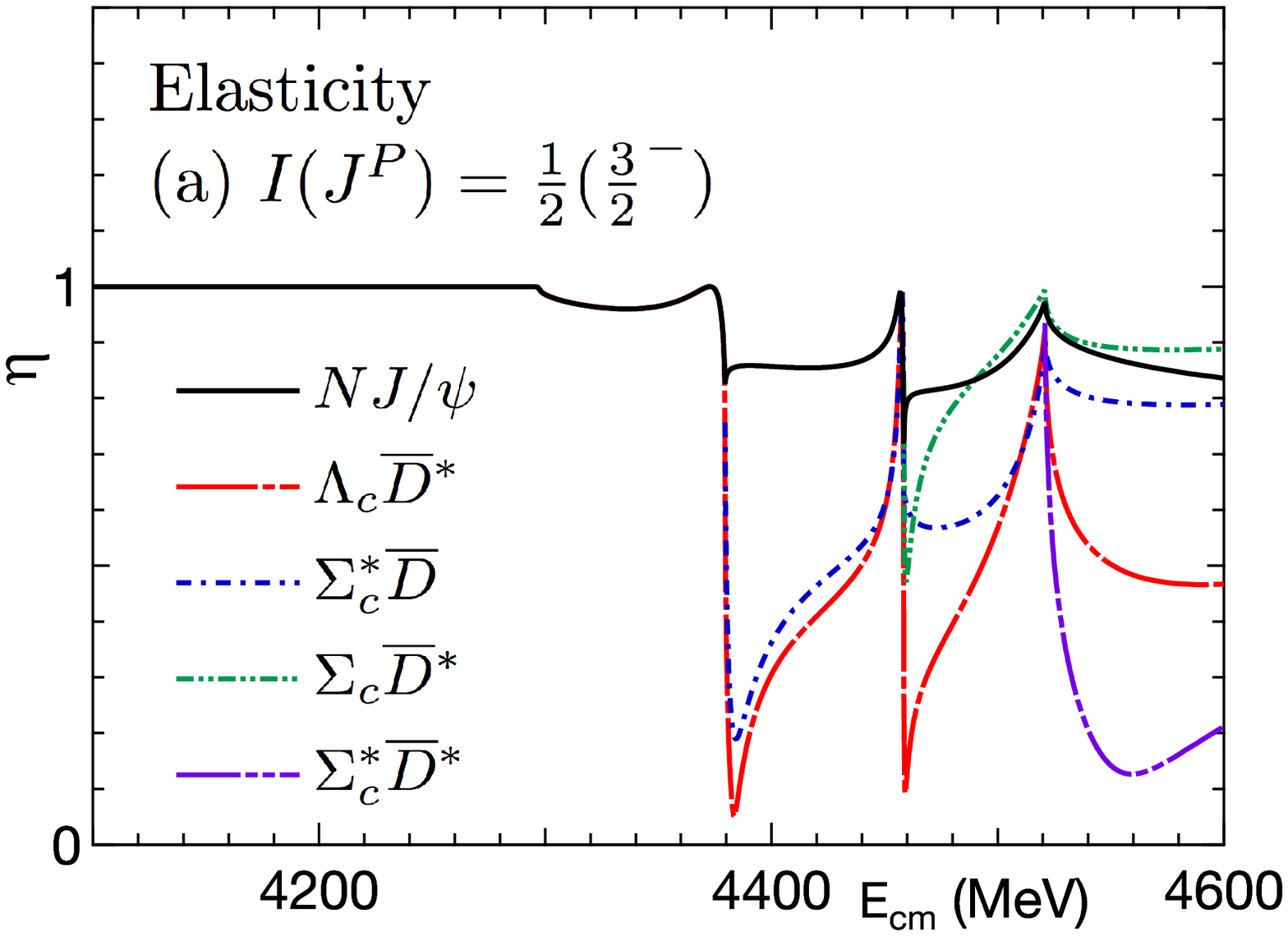}
\includegraphics[width=8cm]{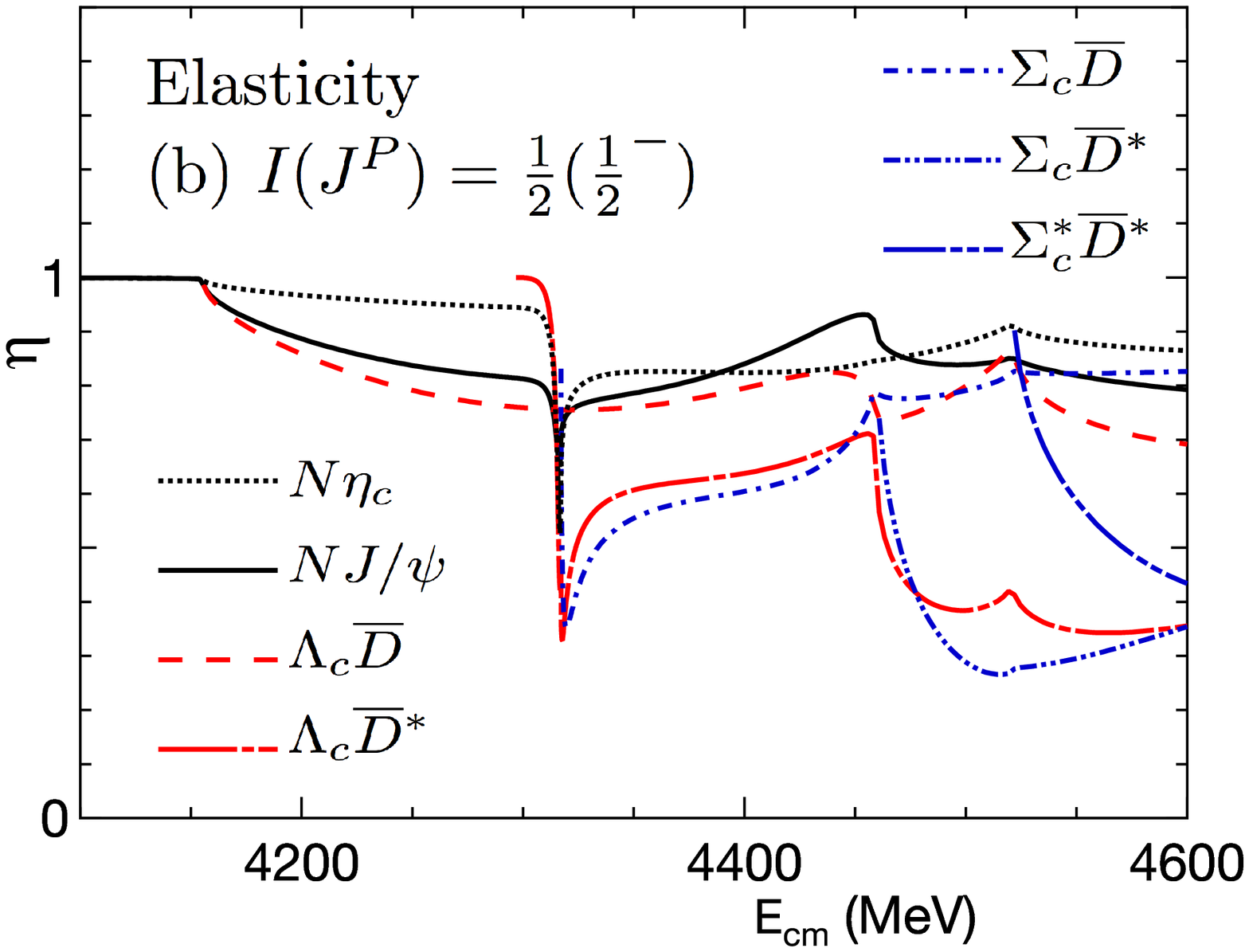}
\caption{The diagonal elasticity of the $S$-wave $uudc\cbar$ $I(J^P)$=$\half1(\half3^-)$  (figure (a)),
and $\half1(\half1^-)$ systems (figure (b)).
For the notation of figures (a) and (b),
see fig.\ \ref{fig:ps23} and
fig.\ \ref{fig:ps21}, respectively.
%In figure (a), the solid line is that of the $N\Jpsi$ channel, 
%the long-dot-dashed, dot-dashed, double-dot-dashed, and  long-double-dot-dashed lines are \Lamc\Dbarstar, \Sigcstar\Dbar, \Sigc\Dbarstar, and \Sigcstar\Dbarstar,
%respectively.
%In figure (b), the solid line is that of the $N\Jpsi$ channel, 
%the dotted, dashed, long-dot-dashed, 
%dot-dashed, double-dot-dashed, and long-double-dot-dashed lines are 
%N$\eta_c$, \Lamc\Dbar, \Lamc\Dbarstar, \Sigc\Dbar, \Sigc\Dbarstar, and \Sigcstar\Dbarstar,
%respectively. (color online)
}
\label{fig:eta}
\end{center}
\end{figure}

There are arguments that
peaks which correspond to the pentaquarks should appear at
$\pi N\rightarrow \Jpsi N$ \cite{Kim:2016cxr}
or
$\gamma N\rightarrow \Jpsi N$ \cite{Wang:2015jsa}
reactions
if 
the coupling between $N\Jpsi$ and the pentaquarks 
is large.
The diagonal elasticities
of the present calculation, $\eta$, which is
the absolute value of the scattering matrix of the same initial and final channels, 
are plotted in Figure \ref{fig:eta}.
In $\half1(\half3^-)$, $\eta$ of the $N\Jpsi$ channel
goes down only 
to 0.81 at the resonance C %of 4379.3 MeV 
or to 0.70 at the cusp B. %of 4458.6 MeV.
The elasticity of the $\half1(\half1^-)$ goes down to 0.65 at the resonance D. %of 4316.5 MeV.
In both of the channels, the mixing between $N\Jpsi$ and the other channels is rather small
except for the resonance energies, 
because the quark overlap is small and 
 because the rearrangement of the charm quark 
is necessary.
The average 
inelasticity of $N\Jpsi$, $1-\eta$, 
over the energy range of the broad  $P_c(4380)$ width
is about 0.1--0.2.
It should be checked whether this value 
is consistent with the above \Jpsi\ production experiments.

%

%\subsection{\Jpsi$N$ scattering length}

The estimate of the cross section $\sigma_{N\Jpsi}$
from the photo production experiment ($\gamma N\rightarrow \Jpsi N$)
is $3.5 \pm 0.8 \pm 0.5$ mb \cite{Anderson:1976hi}.
It corresponds to the scattering length $|a_{N\Jpsi}| = 0.17^{+0.02}_{-0.03}$
fm by using $\sigma=4\pi a^2$.
This scattering length has been 
calculated by many theories.
The QCD sum rule gives 
$a_{N\Jpsi} = -0.10\pm 0.02$ fm
 \cite{Hayashigaki:1998ey}.
The effects of the QCD van der Waals force
was estimated by \cite{Brodsky:1989jd,Brodsky:1997gh},
%which gives 
$a_{N\Jpsi}=-0.24$ fm.
Two quenched Lattice QCD calculations
were reported:
the spin averaged scattering length was obtained as
$-0.71\pm 0.48$ fm or $-0.39\pm 0.14$ fm
\cite{Yokokawa:2006td}, 
or about
$-0.35$ fm (read from figure)\cite{Kawanai:2010ru}.
In the present calculation, the
$\Jpsi N$ scattering length is
$-0.077$ fm for $J$=\half3 and
$-0.103$ fm for $J$=\half1.
The spin averaged value is $a_{N\Jpsi} =-0.085$ fm.
Since the quark interaction does not produce
a direct interaction between 
$N$ and the $c\cbar$ mesons,
this attraction here solely comes from 
the channel coupling between the $N\Jpsi$ and the \Lamc\Dbar$^{(*)}$ and \Sigc$^{(*)}$\Dbar$^{(*)}$
channels.
Our results
suggest that 
roughly half of the observed attraction between the $N$ and \Jpsi\ comes from the channel coupling.

%\section{Misc}

In this work, we employ the quark interaction arising from the gluons.
%which plays an important role in the short range region.
It is because
we would like to investigate the features of the color-octet $uud$ baryons,
which newly appears in the hidden-charm pentaquarks.
As for the \Lamc\Dbar\ or \Sigc\Dbar\ baryon meson channel, however,
it is necessary to include the pion-exchange force in the long range region.
As is reported in ref.\ \cite{Yamaguchi:2016ote},
the meson-exchange models may give many bound states and resonances
though the results seem to depend strongly on the cutoff value.
It is very interesting to see whether the energies of the currently obtained resonances move
or their widths become broader by introducing the meson-exchange in our model.
Moreover, in order to discuss the $P_c(4450)$, which has an opposite parity to the $P_c(4380)$,
simultaneously, 
the $P$-wave baryon-meson relative motion and the positive parity mesons should be introduced.
Since the lowest orbital excitation 
is considered to be the $S$-wave $uud$ configuration with the $P$-wave $c\cbar$ pair,
the color-octet $uud$ configuration may again play an important role there.
We take both of them as future problems.

It is also interesting to investigate the hidden-charm pentaquarks with the strangeness,
where the flavor-octet $uds$ may play a similar role to the present work.
As for the isospin-\half3 $uuuc\cbar$ systems, the three light quarks are either 
 color-octet spin-\half1 
 or  color-singlet spin-\half3.
Because CMI of these light quark configurations
contributes repulsively,
%the values of $\bra\Ocmi\ket_{5q}^{(HQ)}$ 
%are above the \Sigc\Dbar\ threshold,
and because the Pauli-blocking gives strong repulsion there,
 all the five-quark systems 
seem to dissolve by the baryon-meson couplings,
which we will also discuss elsewhere.

The isospin-\half1,
or the flavor-octet hidden-$Q\Qbar$ baryons are a special place to look into for the color-octet $uud$ baryons.
When we introduce the $b\bbar$ pairs instead of $c\cbar$ to the five-quark systems,
the situation will become more manifest.
The system is closer to the heavy quark limit,
and the thresholds of the 
$\Sigma_b^{(*)}B^{(*)}$ are closer to each other.
The four structures we find in the $uudc\cbar$ states are also found to exist.
The bound state in $J^P=\half5^-$ has a binding energy of more than 10 MeV.
We discuss the system with the $b\bbar$ pairs as well as those with 
the $s\overline{s}$ pairs elsewhere.

\section{Conclusions}

The $I(J^P)=\half1(\half1^-)$, $\half1(\half3^-)$, and $\half1(\half5^-)$ $uudc\cbar$ systems 
are investigated by the quark cluster model.
There is no strong repulsion due to the quark Pauli blocking
in the relevant baryon meson systems.
It is shown that the color-octet isospin-\half1 spin-\half3 $uud$ configuration 
gains an attraction from the color magnetic interaction.
The $uudc\cbar$ states with this configuration 
cause structures around the $\Sigc{}^{(*)}\Dbar{}^{(*)}$
thresholds.
We have found
one bound state in $\half1(\half5^-)$, %4520 MeV,
one resonance and a cusp in $\half1(\half3^-)$, and
one resonance in $\half1(\half1^-)$ 
in the negative parity channels.
%We argue that these resonances and cusp may combine to form the broad peak of $P_c(4380)$. 
We argue that one of these structures may give $P_c(4450)$ if its parity is found to be negative, or 
these resonances and cusp may combine to form the broad peak of $P_c(4380)$. 
\bigskip

%\section*{Acknowledgement}
We would like to thank Professors M.\ Oka and
A.\ Hosaka for
useful discussions.


\begin{thebibliography}{99}

%\cite{Aaij:2015tga}
\bibitem{Aaij:2015tga} 
  R.~Aaij {\it et al.} [LHCb Collaboration],
  %``Observation of $J/\psi p$ Resonances Consistent with Pentaquark States in $\Lambda_b^0 \to J/\psi K^- p$ Decays,''
  Phys.\ Rev.\ Lett.\  {\bf 115}, 072001 (2015).
%  doi:10.1103/PhysRevLett.115.072001
%  [arXiv:1507.03414 [hep-ex]].
  %%CITATION = doi:10.1103/PhysRevLett.115.072001;%%
  %200 citations counted in INSPIRE as of 12 Jun 2016

%\cite{Aaij:2016phn}
%\cite{Aaij:2016phn}
\bibitem{Aaij:2016phn} 
  R.~Aaij {\it et al.} [LHCb Collaboration],
  %``Model-independent evidence for $J/\psi p$ contributions to $\Lambda_b^0\to J/\psi p K^-$ decays,''
  Phys.\ Rev.\ Lett.\  {\bf 117}, 082002 (2016).
%  doi:10.1103/PhysRevLett.117.082002
%  [arXiv:1604.05708 [hep-ex]].
  %%CITATION = doi:10.1103/PhysRevLett.117.082002;%%
  %15 citations counted in INSPIRE as of 13 Oct 2016

%%%%%%%%%%%%%%%%%%%%%%  
%%\cite{Huang:2015uda}
%\bibitem{Huang:2015uda} 
%  H.~Huang, C.~Deng, J.~Ping and F.~Wang,
%  %``Possible pentaquarks with heavy quarks,''
%  arXiv:1510.04648 [hep-ph].
%  %%CITATION = ARXIV:1510.04648;%%
%  %15 citations counted in INSPIRE as of 12 Jun 2016
%  
%%\cite{Yang:2015bmv}
%\bibitem{Yang:2015bmv} 
%  G.~Yang and J.~Ping,
%  %``The Structure of Pentaquarks $P_c^+$ in the Chiral Quark Model,''
%  arXiv:1511.09053 [hep-ph].
%  %%CITATION = ARXIV:1511.09053;%%
%  %8 citations counted in INSPIRE as of 12 Jun 2016
%
%%\cite{Anisovich:2015zqa}
%\bibitem{Anisovich:2015zqa} 
%  V.~V.~Anisovich, M.~A.~Matveev, J.~Nyiri, A.~V.~Sarantsev and A.~N.~Semenova,
%  %``Nonstrange and strange pentaquarks with hidden charm,''
%  Int.\ J.\ Mod.\ Phys.\ A {\bf 30}, no. 32, 1550190 (2015)
%%  doi:10.1142/S0217751X15501900
%%  [arXiv:1509.04898 [hep-ph]].
%  %%CITATION = doi:10.1142/S0217751X15501900;%%
%  %11 citations counted in INSPIRE as of 12 Jun 2016

%\cite{Wu:2010jy}
\bibitem{Wu:2010jy} 
  J.~J.~Wu, R.~Molina, E.~Oset and B.~S.~Zou,
  %``Prediction of narrow $N^*$ and $\Lambda^*$ resonances with hidden charm above 4 GeV,''
  Phys.\ Rev.\ Lett.\  {\bf 105}, 232001 (2010).
%  doi:10.1103/PhysRevLett.105.232001
%  [arXiv:1007.0573 [nucl-th]].
  %%CITATION = doi:10.1103/PhysRevLett.105.232001;%%
  %111 citations counted in INSPIRE as of 19 Aug 2016
  
%\cite{Chen:2016qju}
\bibitem{Chen:2016qju} 
  H.~X.~Chen, W.~Chen, X.~Liu and S.~L.~Zhu,
  %``The hidden-charm pentaquark and tetraquark states,''
  Phys.\ Rept.\  {\bf 639}, 1 (2016), and references therein.
%  doi:10.1016/j.physrep.2016.05.004
%  [arXiv:1601.02092 [hep-ph]].
  %%CITATION = doi:10.1016/j.physrep.2016.05.004;%%
  %65 citations counted in INSPIRE as of 11 Aug 2016

%\cite{Oka:2000wj}
\bibitem{Oka:2000wj} 
  M.~Oka, K.~Shimizu and K.~Yazaki,
  %``Quark cluster model of baryon baryon interaction,''
  Prog.\ Theor.\ Phys.\ Suppl.\  {\bf 137}, 1 (2000).
%  doi:10.1143/PTPS.137.1
  %%CITATION = doi:10.1143/PTPS.137.1;%%
  %53 citations counted in INSPIRE as of 02 Aug 2016


%\cite{Takeuchi:2007tv}
\bibitem{Takeuchi:2007tv}
  S.~Takeuchi and K.~Shimizu,
  %``Lambda(1405) as a resonance in the baryon meson scattering coupled to the
  %q**3 state in a quark model,''
  Phys.\ Rev.\  C {\bf 76}, 035204 (2007).
%  [arXiv:0705.0565 [hep-ph]].
  %%CITATION = PHRVA,C76,035204;%%

%\cite{Sasaki:2015ab}
\bibitem{Sasaki:2015ab}  
K.~Sasaki {\it et al.} [HAL QCD Collaboration],
%``Coupled-channel approach to strangeness S=-2 
  %baryon-baryon interactions in lattice QCD,''
Prog. Theor. Exp. Phys. {\bf 2015}, 113B01 (2015).
% doi:10.1093/ptep/ptv144

  
%%%%%%%%%%%%%%%%%%%%%%%%%%%%%
%\cite{Chen:2015loa}
%\bibitem{Chen:2015loa} 
%  R.~Chen, X.~Liu, X.~Q.~Li and S.~L.~Zhu,
%``Identifying exotic hidden-charm pentaquarks,''
%  Phys.\ Rev.\ Lett.\  {\bf 115}, no. 13, 132002 (2015)
%  doi:10.1103/PhysRevLett.115.132002
%  [arXiv:1507.03704 [hep-ph]].
%%CITATION = doi:10.1103/PhysRevLett.115.132002;%%
%56 citations counted in INSPIRE as of 12 Jun 2016
%
%\cite{Roca:2015dva}
%\bibitem{Roca:2015dva} 
%  L.~Roca, J.~Nieves and E.~Oset,
%``LHCb pentaquark as a $\bar{D}^*\Sigma_c-\bar{D}^*\Sigma_c^*$ molecular state,''
%  Phys.\ Rev.\ D {\bf 92}, no. 9, 094003 (2015).
%  doi:10.1103/PhysRevD.92.094003
%  [arXiv:1507.04249 [hep-ph]].
%%CITATION = doi:10.1103/PhysRevD.92.094003;%%
%57 citations counted in INSPIRE as of 12 Jun 2016
%\cite{Roca:2016tdh}
%\bibitem{Roca:2016tdh} 
%  L.~Roca and E.~Oset,
%``On the hidden charm pentaquarks in $\Lambda_b \to J/\psi K^- p$ decay,''
%  arXiv:1602.06791 [hep-ph].
%%CITATION = ARXIV:1602.06791;%%
%5 citations counted in INSPIRE as of 12 Jun 2016
  
  
%\cite{He:2015cea}
%\bibitem{He:2015cea} 
%  J.~He,
%``$\bar{D}\Sigma^*_c$ and $\bar{D}^*\Sigma_c$ interactions and the LHCb hidden-charmed pentaquarks,''
%  Phys.\ Lett.\ B {\bf 753}, 547 (2016).
%  doi:10.1016/j.physletb.2015.12.071
%  [arXiv:1507.05200 [hep-ph]].
%%CITATION = doi:10.1016/j.physletb.2015.12.071;%%
%49 citations counted in INSPIRE as of 12 Jun 2016

%\cite{Kahana:2015tkb}
%\bibitem{Kahana:2015tkb} 
%D.~E.~Kahana and S.~H.~Kahana,
%``LHCb $P_c^+$ Resonances as Molecular States,''
%  arXiv:1512.01902 [hep-ph].
%%CITATION = ARXIV:1512.01902;%%
%3 citations counted in INSPIRE as of 12 Jun 2016

%\cite{Meissner:2015mza}
%\bibitem{Meissner:2015mza} 
%  U.~G.~Meiﾟner and J.~A.~Oller,
%``Testing the $\chi_{c1}\, p$ composite nature of the $P_c(4450)$,''
%  Phys.\ Lett.\ B {\bf 751}, 59 (2015).
%  doi:10.1016/j.physletb.2015.10.015
%  [arXiv:1507.07478 [hep-ph]].
%%CITATION = doi:10.1016/j.physletb.2015.10.015;%%
%43 citations counted in INSPIRE as of 12 Jun 2016

%\cite{Scoccola:2015nia}
%\bibitem{Scoccola:2015nia} 
%  N.~N.~Scoccola, D.~O.~Riska and M.~Rho,
%``Pentaquark candidates P$_c^+$(4380) and P$_c^+$(4450) within the soliton picture of baryons,''
%  Phys.\ Rev.\ D {\bf 92}, no. 5, 051501 (2015).
%  doi:10.1103/PhysRevD.92.051501
%  [arXiv:1508.01172 [hep-ph]].
%%CITATION = doi:10.1103/PhysRevD.92.051501;%%
%29 citations counted in INSPIRE as of 12 Jun 2016
  
%\cite{Wang:2015ava}
%\bibitem{Wang:2015ava} 
%Z.~G.~Wang and T.~Huang,
%``Analysis of the ${\frac{1}{2}}^{\pm }$ pentaquark states in the diquark model with QCD sum rules,''
%  Eur.\ Phys.\ J.\ C {\bf 76}, no. 1, 43 (2016).
%  doi:10.1140/epjc/s10052-016-3880-8
%  [arXiv:1508.04189 [hep-ph]].
%%CITATION = doi:10.1140/epjc/s10052-016-3880-8;%%
%15 citations counted in INSPIRE as of 12 Jun 2016 
  
%\cite{Kopeliovich:2015vqa}
%\bibitem{Kopeliovich:2015vqa} 
% V.~Kopeliovich and I.~Potashnikova,
%``Simple estimates of the hidden beauty pentaquarks masses,''
%  arXiv:1510.05958 [hep-ph].
%%CITATION = ARXIV:1510.05958;%%
%1 citations counted in INSPIRE as of 12 Jun 2016
  
%\cite{Mikhasenko:2015vca}
%\bibitem{Mikhasenko:2015vca} 
%  M.~Mikhasenko,
%``A triangle singularity and the LHCb pentaquarks,''
%  arXiv:1507.06552 [hep-ph].
%%CITATION = ARXIV:1507.06552;%%
%38 citations counted in INSPIRE as of 12 Jun 2016


%\cite{Kawanai:2011xb}
\bibitem{Kawanai:2011xb} 
  T.~Kawanai and S.~Sasaki,
  %``Interquark potential with finite quark mass from lattice QCD,''
  Phys.\ Rev.\ Lett.\  {\bf 107}, 091601 (2011).
%  doi:10.1103/PhysRevLett.107.091601
%  [arXiv:1102.3246 [hep-lat]].
  %%CITATION = doi:10.1103/PhysRevLett.107.091601;%%
  %29 citations counted in INSPIRE as of 30 May 2016
  
  %\cite{Godfrey:1985xj}
\bibitem{Godfrey:1985xj}
  S.~Godfrey and N.~Isgur,
  %``Mesons in a Relativized Quark Model with Chromodynamics,''
  Phys.\ Rev.\ D {\bf 32}, 189 (1985).
  %%CITATION = PHRVA,D32,189;%%
  %1834 citations counted in INSPIRE as of 18 Aug 2013

%\cite{Capstick:1986bm}
\bibitem{Capstick:1986bm} 
  S.~Capstick and N.~Isgur,
  %``Baryons in a Relativized Quark Model with Chromodynamics,''
  Phys.\ Rev.\ D {\bf 34}, 2809 (1986)
  [AIP Conf.\ Proc.\  {\bf 132}, 267 (1985)].
%  doi:10.1103/PhysRevD.34.2809, 10.1063/1.35361
  %%CITATION = doi:10.1103/PhysRevD.34.2809, 10.1063/1.35361;%%
  %999 citations counted in INSPIRE as of 28 Jun 2016

%\cite{Stancu:1996}
\bibitem{Stancu:1996}
See, for example, 
Fl.~Stancu, Group Theory in Subnuclear Physics, 
Oxford University Press, New York, 1996.
%{Stancu}

  
%\cite{Agashe:2014kda}
\bibitem{pdg} 
%\bibitem{Agashe:2014kda} 
  K.~A.~Olive {\it et al.} [Particle Data Group Collaboration],
  %``Review of Particle Physics,''
  Chin.\ Phys.\ C {\bf 38}, 090001 (2014).
%  doi:10.1088/1674-1137/38/9/090001
  %%CITATION = doi:10.1088/1674-1137/38/9/090001;%%
  %4157 citations counted in INSPIRE as of 23 Jun 2016



%\cite{Kim:2016cxr}
\bibitem{Kim:2016cxr} 
  S.~H.~Kim, H.~C.~Kim and A.~Hosaka,
  %``Heavy pentaquark states $P_c(4380)$ and $P_c(4450)$ in the $J/\psi$ production induced by pion beams off the nucleon,''
  arXiv:1605.02919 [hep-ph].
  %%CITATION = ARXIV:1605.02919;%%
  %1 citations counted in INSPIRE as of 02 Aug 2016  
  
%\cite{Wang:2015jsa}
\bibitem{Wang:2015jsa} 
  Q.~Wang, X.~H.~Liu and Q.~Zhao,
  %``Photoproduction of hidden charm pentaquark states $P_c^+(4380)$ and $P_c^+(4450)$,''
  Phys.\ Rev.\ D {\bf 92}, 034022 (2015).
%  doi:10.1103/PhysRevD.92.034022
%  [arXiv:1508.00339 [hep-ph]].
  %%CITATION = doi:10.1103/PhysRevD.92.034022;%%
  %39 citations counted in INSPIRE as of 02 Aug 2016
 
 %\cite{Anderson:1976hi}
\bibitem{Anderson:1976hi} 
  R.~L.~Anderson {\it et al.},
  %``A Measurement of the a-Dependence of psi Photoproduction,''
  Phys.\ Rev.\ Lett.\  {\bf 38}, 263 (1977).
%  doi:10.1103/PhysRevLett.38.263
  %%CITATION = doi:10.1103/PhysRevLett.38.263;%%
  %162 citations counted in INSPIRE as of 05 Jun 2016

%\cite{Hayashigaki:1998ey}
\bibitem{Hayashigaki:1998ey} 
  A.~Hayashigaki,
  %``J / psi nucleon scattering length and in-medium mass shift of J / psi in QCD sum rule analysis,''
  Prog.\ Theor.\ Phys.\  {\bf 101}, 923 (1999).
%  doi:10.1143/PTP.101.923
%  [nucl-th/9811092].
  %%CITATION = doi:10.1143/PTP.101.923;%%
  %46 citations counted in INSPIRE as of 05 Jun 2016

%\cite{Brodsky:1989jd}
\bibitem{Brodsky:1989jd} 
  S.~J.~Brodsky, I.~A.~Schmidt and G.~F.~de Teramond,
  %``Nuclear Bound Quarkonium,''
  Phys.\ Rev.\ Lett.\  {\bf 64}, 1011 (1990).
%  doi:10.1103/PhysRevLett.64.1011
  %%CITATION = doi:10.1103/PhysRevLett.64.1011;%%
  %163 citations counted in INSPIRE as of 01 Jun 2016
  
%\cite{Brodsky:1997gh}
\bibitem{Brodsky:1997gh} 
  S.~J.~Brodsky and G.~A.~Miller,
  %``Is J / psi - nucleon scattering dominated by the gluonic van der Waals interaction?,''
  Phys.\ Lett.\ B {\bf 412}, 125 (1997).
%  doi:10.1016/S0370-2693(97)01045-9
%  [hep-ph/9707382].
  %%CITATION = doi:10.1016/S0370-2693(97)01045-9;%%
  %55 citations counted in INSPIRE as of 01 Jun 2016

%\cite{Yokokawa:2006td}
\bibitem{Yokokawa:2006td} 
  K.~Yokokawa, S.~Sasaki, T.~Hatsuda and A.~Hayashigaki,
  %``First lattice study of low-energy charmonium-hadron interaction,''
  Phys.\ Rev.\ D {\bf 74}, 034504 (2006).
%  doi:10.1103/PhysRevD.74.034504
%  [hep-lat/0605009].
  %%CITATION = doi:10.1103/PhysRevD.74.034504;%%
  %25 citations counted in INSPIRE as of 01 Jun 2016

%\cite{Kawanai:2010ru}
\bibitem{Kawanai:2010ru} 
  T.~Kawanai and S.~Sasaki,
  %``Charmonium-nucleon interaction from lattice QCD with a relativistic heavy quark action,''
  PoS LATTICE {\bf 2010}, 156 (2010).
%  [arXiv:1011.1322 [hep-lat]].
  %%CITATION = ARXIV:1011.1322;%%
  %12 citations counted in INSPIRE as of 05 Jun 2016
  

%%%%%%%%%%%%%%%
%%\cite{Lipkin:1998pb}
%\bibitem{Lipkin:1998pb} 
%  H.~J.~Lipkin,
%  %``Pentaquark update after ten years,''
%  Nucl.\ Phys.\ A {\bf 625}, 207 (1997).
%%  doi:10.1016/S0375-9474(97)81460-1
%%  [hep-ph/9804218].
%  %%CITATION = doi:10.1016/S0375-9474(97)81460-1;%%
%  %33 citations counted in INSPIRE as of 12 Jun 2016
%
%%%%%%%%%%%%%%%
%
%
%%\cite{Kawanai:2010ev}
%\bibitem{Kawanai:2010ev} 
%  T.~Kawanai and S.~Sasaki,
%  %``Charmonium-nucleon potential from lattice QCD,''
%  Phys.\ Rev.\ D {\bf 82}, 091501 (2010).
%%  doi:10.1103/PhysRevD.82.091501
%%  [arXiv:1009.3332 [hep-lat]].
%  %%CITATION = doi:10.1103/PhysRevD.82.091501;%%
%  %27 citations counted in INSPIRE as of 30 May 2016
%
%
%
%
%%\cite{Kharzeev:1996yx}
%\bibitem{Kharzeev:1996yx} 
%  D.~Kharzeev, C.~Lourenco, M.~Nardi and H.~Satz,
%  %``A Quantitative analysis of charmonium suppression in nuclear collisions,''
%  Z.\ Phys.\ C {\bf 74}, 307 (1997).
%%  doi:10.1007/s002880050392
%%  [hep-ph/9612217].
%  %%CITATION = doi:10.1007/s002880050392;%%
%  %276 citations counted in INSPIRE as of 05 Jun 2016
%  
%%\cite{Luke:1992tm}
%\bibitem{Luke:1992tm} 
%  M.~E.~Luke, A.~V.~Manohar and M.~J.~Savage,
%  %``A QCD Calculation of the interaction of quarkonium with nuclei,''
%  Phys.\ Lett.\ B {\bf 288}, 355 (1992).
%%  doi:10.1016/0370-2693(92)91114-O
%%  [hep-ph/9204219].
%  %%CITATION = doi:10.1016/0370-2693(92)91114-O;%%
%  %114 citations counted in INSPIRE as of 01 Jun 2016
%
%
%%%\cite{Kawanai:2010ev}
%%\bibitem{Kawanai:2010ev} 
%%  T.~Kawanai and S.~Sasaki,
%%  %``Charmonium-nucleon potential from lattice QCD,''
%%  Phys.\ Rev.\ D {\bf 82}, 091501 (2010)
%%  doi:10.1103/PhysRevD.82.091501
%%  [arXiv:1009.3332 [hep-lat]].
%%  %%CITATION = doi:10.1103/PhysRevD.82.091501;%%
%%  %27 citations counted in INSPIRE as of 05 Jun 2016
%
%  
%%\cite{Kawanai:2015tga}
%\bibitem{Kawanai:2015tga} 
%  T.~Kawanai and S.~Sasaki,
%  %``Potential description of charmonium and charmed-strange mesons from lattice QCD,''
%  Phys.\ Rev.\ D {\bf 92}, no. 9, 094503 (2015).
%%  doi:10.1103/PhysRevD.92.094503
%%  [arXiv:1508.02178 [hep-lat]].
%  %%CITATION = doi:10.1103/PhysRevD.92.094503;%%
%  %4 citations counted in INSPIRE as of 05 Jun 2016
%
%%\cite{Kawanai:2011zz}
%\bibitem{Kawanai:2011zz} 
%  T.~Kawanai and S.~Sasaki,
%  %``Charmonium-nucleon interaction from lattice QCD with 2+1 flavors of dynamical quarks,''
%  AIP Conf.\ Proc.\  {\bf 1388}, 640 (2011).
%%  doi:10.1063/1.3647474
%  %%CITATION = doi:10.1063/1.3647474;%%
%
%%\cite{Liu:2001yx}
%\bibitem{Liu:2001yx} 
%  W.~Liu, C.~M.~Ko and Z.~W.~Lin,
%  %``J / psi absorption cross-section by nucleon,''
%  nucl-th/0107058.
%  %%CITATION = NUCL-TH/0107058;%%
  %6 citations counted in INSPIRE as of 05 Jun 2016

%\cite{Yamaguchi:2016ote}
\bibitem{Yamaguchi:2016ote} 
  Y.~Yamaguchi and E.~Santopinto,
  %``Hidden-charm pentaquarks as a meson-baryon molecule with coupled channels for $\bar{D}^{(\ast)}\Lambda_{\rm c}$ and $\bar{D}^{(\ast)}\Sigma^{(\ast)}_{\rm c}$,''
  arXiv:1606.08330 [hep-ph].
  %%CITATION = ARXIV:1606.08330;%%
  
%%\cite{Yoshida:2015tia}
%\bibitem{Yoshida:2015tia} 
%  T.~Yoshida, E.~Hiyama, A.~Hosaka, M.~Oka and K.~Sadato,
%  %``Spectrum of heavy baryons in the quark model,''
%  Phys.\ Rev.\ D {\bf 92}, no. 11, 114029 (2015).
%%  doi:10.1103/PhysRevD.92.114029
%%  [arXiv:1510.01067 [hep-ph]].
%  %%CITATION = doi:10.1103/PhysRevD.92.114029;%%
%  %7 citations counted in INSPIRE as of 30 May 2016

%\bibitem{Uchino} 
%T.~Uchino, W.~H.~Liang and E.~Oset,
%``Baryon states with hidden charm in the extended local hidden gauge approach,''
% Eur.\ Phys.\ J.\ A {\bf 52}, no. 3, 43 (2016)

\end{thebibliography}
\end{document}